\documentclass{ws-procs9x6}

\def\etal {{\it et al.}}

\begin{document}

\title{LIMITS ON LORENTZ VIOLATION IN\\
NEUTRAL-KAON DECAY}

\author{K.K.\ VOS,$^*$ H.W.\ WILSCHUT and R.G.E.\ TIMMERMANS}

\address{Kernfysisch Versneller Instituut, University of Groningen\\
9747 AA Groningen, The Netherlands\\
$^*$E-mail: K.K.Vos@rug.nl}

\begin{abstract}
The KLOE collaboration recently reported bounds on the directional dependence of the lifetime of the short-lived neutral kaon $K_S$ with respect to the cosmic microwave background dipole anisotropy. We interpret their results in a general framework developed to probe Lorentz violation in the weak interaction. In this approach a Lorentz-violating tensor $\chi_{\mu\nu}$ is added to the standard propagator of the $W$ boson. We derive the $K_S$ decay rate in a naive tree-level model and calculate the asymmetry for the lifetime. By using the KLOE data the real vector part of $\chi_{\mu\nu}$ is found to be smaller than $10^{-2}$. We briefly discuss the theoretical challenges concerning nonleptonic decays. 
\end{abstract}

\bodymatter

\section{Testing Lorentz violation in the weak interaction}

Recently Lorentz violation in the weak interaction has been studied in allowed \cite{Jacob1, Hans1} and forbidden \cite{Jacob2} $\beta$-decay. A general theoretical framework was developed for these tests \cite{Jacob1}, in which the standard $W$-boson propagator is modified by a general Lorentz-violating tensor $\chi_{\mu\nu}$. In the low-energy limit the modified propagator is
\begin{equation}\label{eq:Wboson}
\left\langle W^{\mu+}(q)W^{\nu-}(-q)\right\rangle = -i\,\frac{g^{\mu\nu}+\chi^{\mu\nu}}{M_W^2} \ .
\end{equation}
To constrain $\chi_{\mu\nu}$ we derive the $K_S$ decay rate with Lorentz violation and use data from the KLOE collaboration \cite{KLOE}, which searched for the directional dependence of the lifetime of $K_S$ mesons. 

\section{KLOE lifetime asymmetry}

The KLOE collaboration measured the lifetime asymmetry of the neutral $K_S$ meson, by transforming the $K_S$ momenta event-by-event to galactic coordinates $\left\{\ell,b\right\}$, where $\ell$ is the galactic longitude and $b$ is the galactic latitude. The lifetime asymmetry is defined as
\begin{equation}
\mathcal{A} =\frac{\tau^+ - \tau^-}{\tau^++\tau^-} ,
\end{equation}
where $\tau^{+(-)}$ is the lifetime parallel (antiparallel) to a specific direction in space. Three directions were studied: CMB0, the direction of the dipole anisotropy in the Cosmic Microwave Background (CMB),
and two perpendicular directions, CMB1 and CMB2. 
Only events inside a cone of $30^\circ$ opening angle were used. The measured asymmetries are given in Table \ref{table:life}, where the errors are mainly statistical.

\begin{table}
\tbl{Observed $K_S$ lifetime asymmetry.}
{\begin{tabular}{@{}lcc@{}}\toprule
$\left\{\ell,b\right\}$ & $\mathcal{A}\times 10^3$  & Ref. \\
\colrule
CMB0 = $\left\{264^\circ, 48^\circ\right\}$
 & $-0.2\pm1.0$  & \refcite{KLOE}\\
CMB0 = $\left\{264^\circ, 48^\circ\right\}$ & $-0.13\pm0.4$  & \refcite{dirvio}\\
CMB1 = $\left\{174^\circ, 0^\circ\right\}$ 
 & $0.2\pm1.0$ & \refcite{KLOE}\\
CMB2 = $\left\{264^\circ, -42^\circ\right\}$
 & $0.0\pm0.9$ & \refcite{KLOE} \\
 \botrule
\end{tabular}}
\label{table:life}

\end{table}

\section{$\Delta I=1/2$ rule}\label{sec:1/2}

\begin{figure}[htpb]
	\centering
		\includegraphics[width=0.50\textwidth]{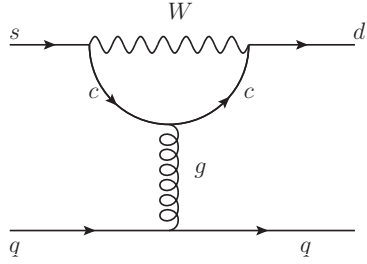}
	\caption{The penguin diagram for an $s$ to $d$ quark transition}
	\label{fig:sdpenguin}
\end{figure}

The description of nonleptonic processes is somewhat more involved than leptonic and semileptonic processes, due to the importance of gluon loops. These QCD corrections can be of the same order as tree-level contributions. Within the Standard Model,
nonleptonic $\Delta S=1$ decays are usually described by an effective lagrangian \cite{leffart}, which contains different operators and their coefficients. An important consequence of the gluon loops is that they generate a so-called penguin diagram, depicted in Fig.\ \ref{fig:sdpenguin}, which can effectively be described by an operator that also contains right-handed quarks. 
For $K_S$ decays there are two isospin $I$ final states, $I=0$ and $I=2$. Experimentally it is found that the former, a $\Delta I=1/2$ transition, is enhanced compared to the latter, a $\Delta I=3/2$ transition. This enhancement is an order of magnitude larger than expected from theoretical calculations with an effective lagrangian, and is known as the $\Delta I=1/2$ rule. The penguin diagram generates a $\Delta I =1/2$ operator $\mathcal{O}_5$ that explains part of the enhancement due to the coupling to right-handed quarks, but the enhancement is not sufficient to explain the $\Delta I=1/2$ rule \cite{donpen}.    

The derivation of the complete Lorentz violating lagrangian is more involved, due to mixing between operators, and lies beyond the scope of this work. The penguin diagram does not mix and we are able to derive the Lorentz violating operator associated with this diagram by integrating out the $W$ boson, following Ref.\ \refcite{wisewitten}. We find \cite{Keri1},
\begin{equation}
\mathcal{O}_5^{LV+SM} = -\frac{1}{2} \bar{d}_L t^a \left[-2g_{\mu\nu}+\chi_{\mu\nu}+\chi_{\nu\mu} + \tilde{\chi}_{\mu\nu} \right] \gamma^{\nu} s_L \left( \bar{q}_R t^a \gamma^{\mu} q_R\right) + \textrm{h.c.},
\end{equation}
where the $g_{\mu\nu}$ part is the Standard Model part and $\tilde{\chi}_{\mu\nu}\equiv i \epsilon^{\alpha\beta\mu\nu}\chi_{\alpha\beta}$. Further discussion and implications of $\mathcal{O}_5$ are given in Ref.\ \refcite{Keri1}. The outstanding problem of the $\Delta I =1/2$ rule makes our calculation with Lorentz violation challenging. We therefore first explore the possibilities of the KLOE data with a naive tree-level model.   

\section{Tree-level model}

To explore the possible bounds from the $K_S$ decay we first use a tree-level model, in which we assume that the dominant process for $K_S$ decay is the tree-level decay. We thus treat this process naively as being semileptonic. 
For the tree-level amplitude with the modified $W$-boson propagator we find
\begin{equation}
\mathcal{M} \sim \left\langle \pi^+|\bar{u}_L\gamma^{\mu}d_L|0\right\rangle (g_{\mu\nu}+\chi^*_{\mu\nu})\left\langle \pi^-|\bar{s}_L\gamma^{\nu}u_L|K^0\right\rangle . 
\end{equation}
After integrating, we find for the asymmetry,
\begin{equation}
\mathcal{A}_{\vec{n}}  = -\frac{\frac{4}{3}+\frac{2}{3}\frac{m_{\pi}^2}{m_K^2}}{(1-\beta_K^2)\left(1-\frac{m_{\pi}^2}{m_K^2}\right)} \,(\chi^r_{i0}+\chi^r_{0i})\beta_K^i = -0.343 (\chi^r_{i0}+\chi^r_{0i})\,\hat{\beta}_K^i ,
\end{equation}
where $2\chi_{\mu\nu}^r=\chi_{\mu\nu}+\chi_{\mu\nu}^*$ and $\hat{\beta}_K$ is the normalized kaon velocity, where for the last part we used $\beta=0.2$. 
The asymmetry thus gets a $\gamma^2$ enhancement and is sensitive to the symmetric real vector part of $\chi_{\mu\nu}$. 

For our final results we transform $\chi_{\mu\nu}$ to the Sun-centered reference frame\cite{datatables} and find at 95\% confidence level\cite{Keri1},
\begin{eqnarray}
 |X^r_{10}+X^r_{01}| & < & 3 \times 10^{-3} , \nonumber \\
 |X^r_{20}+X^r_{02}| & < & 6 \times 10^{-3}  , \nonumber  \\
 |X^r_{30}+X^r_{03}| & < & 6 \times 10^{-3}  . 
\end{eqnarray}

\section{Conclusion and outlook}

We investigated Lorentz violation in neutral $K_S$ decays by modifying the $W$-boson propagator. This nonleptonic decay is theoretically more challenging than semileptonic or leptonic decays. Using a tree-level model we are able to put bounds on the real vector part of $\chi_{\mu\nu}$ of the order of $10^{-2}$. These bounds complement bounds from allowed\cite{Hans1} and forbidden\cite{Jacob2} $\beta$-decay, but are several orders of magnitude less strict than bounds in other sectors\cite{datatables}. In order to significantly improve these bounds a possibility would be to use the $\gamma^2$ enhancement that occurs in decay asymmetries. From the theoretical point of view we argued that nonleptonic decays are much more involved due to QCD corrections, making experiments with leptonic or semileptonic decays preferable.   

\section*{Acknowledgments}

This research was supported by the Dutch Stichting voor Fundamenteel Onderzoek der Materie
(FOM) under Programmes 104 and 114 and project 08PR2636.

\end{document}